\documentclass[pra,twocolumn,showpacs,letterpaper,showpacs,superscriptaddress]{revtex4}
\usepackage{graphicx,amsmath,amssymb,amsfonts,latexsym,color,dcolumn,bm}

\begin{document}

\newcommand{\beq}{\begin{equation}}
\newcommand{\eeq}{  \end{equation}}
\newcommand{\bea}{\begin{eqnarray}}
\newcommand{\eea}{  \end{eqnarray}}
\newcommand{\bit}{\begin{itemize}}
\newcommand{\eit}{  \end{itemize}}

\title{Quantum metrology at the Heisenberg limit with ion traps} 

\author{D.A.R. Dalvit}
\affiliation{Theoretical Division, MS B213, Los Alamos National Laboratory, 
Los Alamos, NM 87545, USA}

\author{R. L. de Matos Filho} 
\affiliation{Instituto de F\'{\i}sica,
Universidade Federal do Rio de Janeiro, Caixa Postal 68.528, 21.941-972, Rio
de Janeiro, Brazil}

\author{F. Toscano} 
\affiliation{Funda\c c\~ao Centro de
Ci\^encias e Educa\c c\~ao Superior \`a Dist\^ancia do Estado do Rio de
Janeiro, 20943-001 Rio de Janeiro, RJ, Brazil}

\affiliation{Instituto de F\'{\i}sica,
Universidade Federal do Rio de Janeiro, Caixa Postal 68.528, 21.941-972, Rio
de Janeiro, Brazil}
  
\date{\today} 

\begin{abstract} 
Sub-Planck phase-space structures in the Wigner function of the motional
degree of freedom of a trapped ion can be used to perform weak force
measurements with Heisenberg-limited sensitivity. We propose methods to
engineer the Hamiltonian of the trapped ion to generate states with such
small scale structures, and we show how to use them in quantum metrology
applications.
\end{abstract} 


\maketitle 


\section{Introduction}

The determination of small parameters has recently acquired a substantial
improvement through quantum measurement, as it is now known that using probes
prepared in judiciously chosen quantum states can increase their sensitivity to perturbations.
As a consequence, quantum metrology has become a subject of great practical
interest \cite{Giovannetti2004}.  The estimation of an unknown parameter of a
quantum system typically involves a three-step process: the initial
preparation of a probe in a known quantum state, the interaction between the
probe and the system to be measured, and a final read-out stage, where the
state of the probe is determined. Typical situations are those when the system
imprints an unknown parameter $x$ onto the probe through a unitary
perturbation $\hat{U}_x=\exp(-i\,x\,\hat{G})$, where the generator $\hat{G}$
is a known Hermitian operator.  The unknown small parameter $x$ of the system
can be inferred by comparing the input and output states of the probe. This
framework captures two important tasks in quantum metrology: i) 
high precision phase measurements $x=\theta$, where $\hat{U}_\theta$ generates
a rotation in phase space of the quantum state of the probe around the origin,
and ii) detection of weak forces that induce a linear displacement $x=s$
of the quantum state of the probe in some direction of phase space.

The accuracy of the parameter estimation is limited by the physical resources
involved in the measurement. Techniques involving probes prepared in
quasiclassical states, such as coherent states of light, have sensitivities at
the standard quantum limit (SQL), also known as the shot-noise limit in the
phase detection situations.  Indeed, in the usual dimensionless phase-space
used in quantum optics the Wigner distribution of a coherent state, with
$\overline{n}$ mean number of photons, is centered at a distance $\simeq
\sqrt{\bar n}$ from the origin, with a width $1/2$.  Thus, the associated
input and output states are distinguishable (approximately orthogonal) when
their respective phase-space distributions are displaced by a minimal distance
$1/2\simeq {\cal O}(\overline{n}^0)$ from one another, i.e., the SQL for
weak displacement measurement does not depend on the number $\overline{n}$ of
photons involved in the measurement. For phase detection the
smallest noticeable rotation occurs when the centers of the phase-space
distributions of the input and output states have an angular distance $(1/2)/
\sqrt{\bar n}$ measured from the origin, i.e., the SQL for phase
measurement scales as $1/\sqrt{\bar n}$.

Using the same physical resources in addition to quantum effects, such as
entanglement or squeezing, sub SQL precision can be achieved
\cite{Caves1981,Yurke1986,Holland1993,Bollinger1996,
Sanders1995,Munro2002,Pezze2006}.
For example,
probes prepared in quantum superpositions of, say, coherent states of light,
have Wigner distributions with sub-Planck phase-space structures of typical
linear size $1/\sqrt{\overline{n}}$ \cite{Zurek2001}.  It was recently shown
\cite{Toscano2006} how these structures can be used to achieve
Heisenberg-limited sensitivity in weak force and phase measurements: the
approximate orthogonality that allows to distinguish the input and output
quantum states of the probe occurs when the peaks (valleys) of the sub-Planck
structures of one come on top of the valleys (peaks) of the sub-Planck
structures of the other. The minimal linear displacement required for this
destructive interference is $\simeq 1/\sqrt{\overline{n}}$, defining the 
so-called Heisenberg limit (HL) for weak force detection.  For rotations the
sub-Planck phase-space structures of the Wigner function of the input state
have to be at a typical distance $\simeq \sqrt{\overline{n}}$ from the origin,
thus the minimal rotation angle is $\simeq 1/\overline{n}$, that defines the
HL scale for phase detection. Hence, the linear size of the sub-Planck
structures sets the sensitivity limit on a probe and states that saturate the
limit on the smallest phase-space structures can allow one to attain
Heisenberg-limited sensitivity.  Sub SQL sensitivities, approaching the
ultimate Heisenberg limit, have been recently achieved experimentally using
multi-qubit states formed by entangled internal degrees of freedom of photons
(polarization entanglement)
\cite{Zeilinger2004,Steinberg2004,Zhao2004,Eisenberg2005}, and of trapped ions
(spin entanglement) \cite{Sackett2000,Meyer2001,Liebfried2004}.

Sub-SQL precision measurements can also be achieved with continuous variables, 
such as the amplitude/phase of a photon trapped in a QED cavity, or the
center of mass motion of trapped ions \cite{Gilchrist2004}.  Such spatial
modes, that can be approximately described as harmonic oscillators, can be prepared in
nonclassical quantum states, e.g.  superpositions of $M$ coherent states, that
possess sub-Planck phase-space structures.  In \cite{Toscano2006} it was shown
how to use them for Heisenberg-limited precision measurements of weak forces
and phase measurement, by entangling the oscillator with a two-level system.
It was also proposed a concrete implementation of this strategy in the context
of cavity QED (that can be extrapolated to trapped ions): a two-level atom
gets entangled with the cavity mode through the (resonant or non-resonant)
Jaynes-Cummings interaction in such a way that, when the atom reaches the
center of the cavity, the photon field is a quantum superposition of two
coherent states ($M=2$).  Such ``cat state" has sub-Planck oscillations
parallel to the line joining the two coherent states, and is therefore
Heisenberg-limited sensitive to perturbations that induce displacements that
are perpendicular to that line.  After the perturbation is applied, the
Jaynes-Cummings evolution is undone, and the populations of the two-level atom
are measured once it leaves the cavity.  This provides information about the
perturbation, that can be estimated with HL precision.  An analysis of the
decoherence processes \cite{Toscano2006} that may affect this scheme shows
that this proposal should be within reach of cavity QED and ion trap
experiments.

One disadvantage of the $M=2$ states is that their sensitivity gradually
degrades as the direction of the perturbing force moves away from the
direction orthogonal to the line joining the two coherent states. Higher order
($M>2$) superpositions of coherent states on a circle do not suffer from these
limitations. In the context of cavity QED, there have been a number of schemes
proposed to generate such states. In some of these proposals
\cite{Brune1992,Parkins1993,Garraway1994,Szabo1996,Guo1996} it is necessary to
send a sequence of atoms through the cavity and perform conditional
measurements (post-selection), limiting the probability of succeeding in
engineering the cavity in the desired coherent state superpositions.  There
are also schemes that involve only one atom, and make use of a single-atom
interferometric method \cite{Domokos1994}, or a dispersive interaction between
the atom and the cavity field \cite{Zeng1998-1}. The first scheme has the
disadvantage of requiring post-selection, while the second one needs very long
interaction times.  Similar strategies have been proposed in the literature in
the context of trapped ions \cite{Zeng1998-2,Jose2000}.

In this paper we propose ways for deterministic generation of higher order
($M>2$) coherent state superpositions with trapped ions, and how to use them
to measure weak forces at the Heisenberg limit.  These states do not suffer
from the limitations of the $M=2$ states generated via the linear
Jaynes-Cummings interactions, as discussed above. In particular, we will
concentrate on how to prepare probes with $M=4$ (``compass state") \cite{Zurek2001}
and how to use them to measure small displacements (weak force detection). Our setups
can also be used to measure small rotations in phase space (phase detection)
by simply adding an appropriate displacement before the application of the
perturbation, as described in \cite{Toscano2006}.

The paper is organized as follows. In Section II we discuss the sensitivity to
perturbations of the $M=4$ compass state and we review the strategy for weak
force measurements.  In Section III we review the engineering of the ion-laser
interaction for a single trapped ion. Section IV contains two possible
approaches for generating compass states: the first one uses conditional
quantum gates between the internal and the motional degrees of freedom of the ion,
and the second one uses a nonlinear ion-laser engineered interaction. In
Section V we implement the scheme for measuring weak forces with
Heisenberg-limited sensitivity.  Finally, Section VI summarizes our central
results.


\section{Heisenberg-limited quantum metrology with continuous variables}

In this section we review how to perform  continuous variable based quantum metrology at the Heisenberg limit by using sub-Planck phase-space structures 
present in circular coherent states of a harmonic oscillator (the probe)
\cite{Toscano2006}.  In the following we will focus on perturbations that
induce a small phase-space linear displacement of the quantum state of the
probe, for example when a small force is exerted onto the probe.  The unitary
operator describing such a displacement is 
$\hat{D}(\beta)= e^{\beta\hat{a}^{\dagger}-\beta^*\hat{a}}$, 
where $\beta$ is an arbitrary
small displacement in phase-space with a magnitude $|\beta| \ll |\alpha|$. In
this approximation the unitary operation $\hat{D}(\beta)$ takes any coherent
state $|\alpha\rangle$ to $e^{i {\rm Im}(\beta \alpha^*)} |\alpha+\beta
\rangle \approx e^{2 i {\rm Im} (\beta \alpha^*)} |\alpha\rangle$.
Perturbations inducing small rotations $\hat{R}(\theta)=e^{i \theta
\hat{a}^{\dagger} \hat{a}} (\theta \ll 1)$ in phase space can be treated in
a similar way, as described in \cite{Toscano2006}.

\begin{figure*}[t]
\setlength{\unitlength}{1cm}
\begin{center}
\scalebox{0.95}[0.95]{%
\includegraphics*[width=6.2cm,angle=-90]{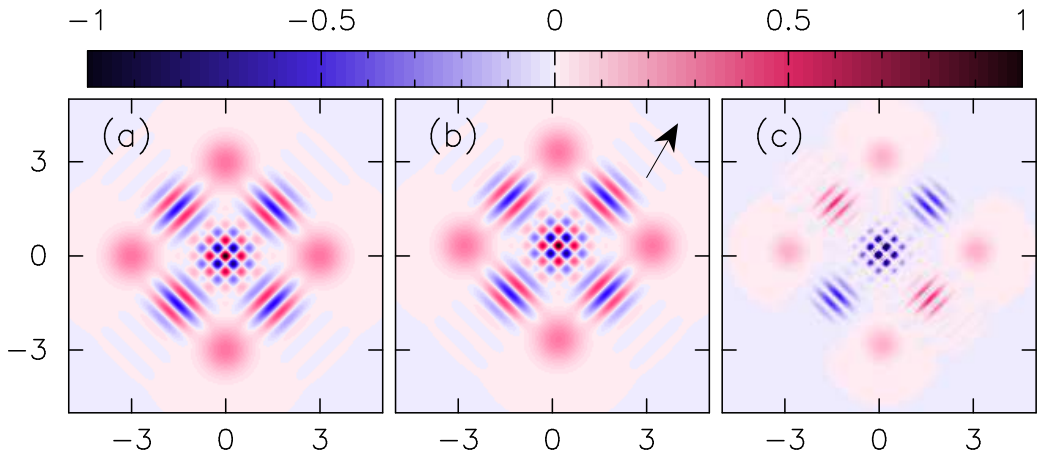}}
\end{center}
\caption{(Color) The Wigner functions in the $\alpha$ plane for: (a) the compass state
$|\mbox{cat}_4\rangle$ with $\alpha=3$, and (b) the displaced compass state
$\hat{D}(\beta)|\mbox{cat}_4\rangle=|\mbox{cat}_4(s_o)\rangle$ where
$\beta=e^{i\varphi}s_o\alpha/|\alpha|$ is a displacement of magnitude
$|\beta|=s_o=\pi/ 2\,b_+(\varphi)|\alpha|$ in the direction relative to
$\alpha$ given by $\varphi=\pi/3$.  Here $b_+(\varphi)=\cos(\varphi)+\sin(\varphi)$.
The black arrow in (b) indicates the
direction of the displacement $\beta$.  In (c) we display the product of the
unpertubed and the perturbed Wigner functions.  When performing the
integration over the $\alpha$ plane of this product of Wigner functions,
that is equal to the overlap $|\langle\mbox{cat}_4|\mbox{cat}_4(s_o)\rangle|^2$, 
the negative contributions (in blue) cancel the positive ones (in red), leading to
quasiorthogonality.}
\label{fig1}
\end{figure*}

Circular coherent states are a special kind of states of harmonic
oscillators, formed by the superposition of $M$ coherent states equidistantly
placed on a circle $\cal{C}$ of radius $|\alpha|$.  They are defined as
\begin{equation}
\label{defcatM}
| {\rm cat}_M \rangle = \frac{{\cal N}}{\sqrt{M}} \sum_{k=1}^M e^{i \gamma_k} 
\; | e^{i \varphi_k} \alpha \rangle ,
\end{equation}
where $\varphi_k=2 \pi k / M$, $\gamma_k$ are arbitrary phases, and 
${\cal N}=1+{\cal O}(e^{-|\alpha|^2})$ is a normalization constant.  When
$|\alpha|\gtrsim 3$, it is a good approximation to take ${\cal N}=1$, and we
will assume this is the case in the following considerations.  These 
non-classical states of a harmonic oscillator have a mean number of
excitations $\bar{n}\equiv\langle {\rm cat}_M|\hat{a}^{\dagger}\hat{a}|{\rm
cat}_M\rangle \approx |\alpha|^2$. Examples of these states are the
Schr\"odinger cat state ($M=2$), considered in detail in \cite{Toscano2006}
for weak force and phase measurements, and the compass state
\begin{equation}
|\mbox{cat}_4\rangle\equiv
\frac{1}{2}\left(e^{i\gamma_1}|i\alpha\rangle+
e^{i\gamma_2}|-\alpha\rangle+
e^{i\gamma_3}|-i\alpha\rangle+
e^{i\gamma_4}|\alpha\rangle\right) ,
\label{defcat4}
\end{equation}
that we consider here. 

\begin{figure}[t]
\setlength{\unitlength}{1cm}
\begin{center}
\scalebox{0.95}[0.95]{%
\includegraphics*[width=6.2cm,angle=-90]{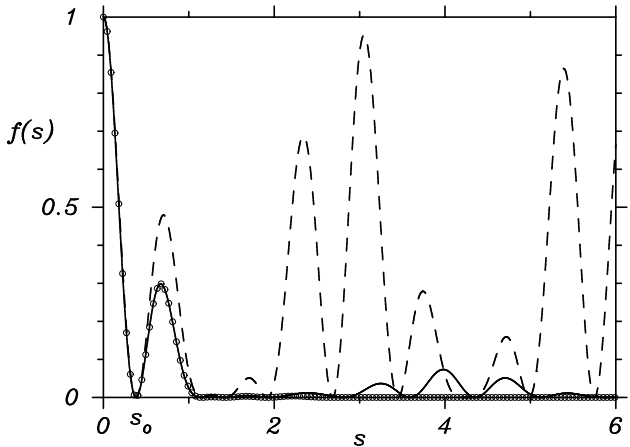}}
\end{center}
\caption{The fidelity $f(s)\equiv|\langle \mbox{cat}_4|\mbox{cat}_4(s)\rangle|^2$
as a function of the magnitude of the linear displacement $s=|\beta|$ (full
line) where $\beta=e^{i\varphi}s\alpha/|\alpha|$ with $\varphi=\pi/3$.  The
circles $(\circ)$ correspond to points given by the approximation in
Eq.(\ref{fidelity1-cat4}).  The dashed line is the approximation in
Eq.(\ref{fidelity2-cat4}) The displacement corresponding to quasiorthogonality, i.e.,
$s_o$ in Eq.(\ref{def-so}) is indicated in the plot.}
\label{fig2}
\end{figure}

Figure \ref{fig1} depicts the Wigner function of the
compass state and shows the sub-Planck phase-space structures, that oscillate
with a typical wavelength $\sim 1/|\alpha|$. These structures are responsible
for the Heisenberg-limited sensitivity of these  states for quantum
metrology applications. Indeed, the smallest linear displacement needed to
distinguish the unperturbed $|\mbox{cat}_4\rangle$ from the perturbed
$|\mbox{cat}_4(s)\rangle \equiv \hat{D}(\beta)|\mbox{cat}_4\rangle$ state of
the probe (and, therefore, to attain quasi-orthogonality) is $s\sim
1/|\alpha|$, for any direction $\beta=e^{i\varphi}s\alpha/|\alpha|$. As it was
shown in \cite{Toscano2006}, the sub-Planck structures of the Wigner functions
of the circular states determine the oscillatory behavior of the fidelity
function $f(s) \equiv|\langle \mbox{cat}_4|\mbox{cat}_4(s)\rangle|^2$, that
for the compass state reads
\begin{eqnarray}
f(s)&\approx& \frac{1}{16}\left|\sum_{k=1}^{4} \;
e^{i {\rm Im} (\beta \alpha_k^*)}\;
\langle \alpha_k|\alpha_k+\beta\rangle \right|^2 \nonumber\\
&=&
e^{-s^2}\,
\cos^2\big(b_{+}(\varphi)|\alpha|s\big)
\cos^2\big(b_{-}(\varphi)|\alpha|s\big) ,
\label{fidelity1-cat4} 
\end{eqnarray}
where $\alpha_k\equiv e^{i \varphi_k}\alpha$ and
$b_{\pm}(\varphi)\equiv \cos(\varphi)\pm\sin(\varphi)$
($\varphi$ is the angle between $\beta$ and $\alpha$). 
Here we have neglected contributions 
$\langle \alpha_l|\alpha_k+\beta\rangle\approx {\cal O}\big(e^{-|\alpha|^2}\big)$ 
for $l\neq k$, that is a good approximation for $|\alpha|\gtrsim 3$, and we have
used that $\langle \alpha_k|\alpha_k+\beta\rangle=e^{-|\beta|^2}
\exp\left\{i{\rm Im}(\beta \alpha_k^*)\right\}$.
When $e^{-s^2}\approx 1$, valid for $|\beta|=s\ll 1$, we
can re-write the fidelity function as 
\begin{equation} 
f(s) \approx
\cos^2\big(b_{+}(\varphi)|\alpha|s\big)
\cos^2\big(b_{-}(\varphi)|\alpha|s\big) .
\label{fidelity2-cat4}
\end{equation}
Note that $f(s)$ does not depend on the phases
$\gamma_k$ that enter in the definition of the compass state
$|\mbox{cat}_4\rangle$ in Eq.(\ref{defcat4}).
We can see in Figure \ref{fig2} the oscillatory behavior
of the fidelity $f(s)$ as a function of the magnitude of the linear
displacement $s=|\beta|$ with a typical wavelength $\sim 1/|\alpha|$.
The minimal displacement to achieve quasi-orthogonality, $s_o$, can be 
obtained equating to zero the approximation in Eq.(\ref{fidelity2-cat4}) because
we see in Fig.\ref{fig2} that this approximation well
describes the fidelity beyond $s_o$. 
Thus, the minimal displacement to achieve quasi-orthogonality is approximately 
given by
\begin{equation}
\label{def-so}
s_o\approx \frac{\pi}{2b(\varphi)|\alpha|}, 
\end{equation}
where $b(\varphi)\equiv\mbox{max}\{|b_+(\varphi)|,|b_-(\varphi)|\}$ (see
Fig.(\ref{fig2})).    The scaling of $s_o$ 
as $1/|\alpha|$ implies that one can detect
displacement perturbations with Heisenberg-limited sensitivity.  Note that
$b_+(\varphi)$ and $b_-(\varphi)$ are never simultaneously zero in the
interval $[0,2\pi]$, which means that for any direction of the displacement
$\beta$ there is a minimum (finite) displacement for which the unperturbed and
the perturbed compass states are quasi-orthogonal.  Contrary to the cat state
($M=2$), whose HL sensitivity degrades as the direction of the perturbation
$\beta$ tends to the direction of $\alpha$ (i.e., as $\varphi \rightarrow 0$),
compass states are HL sensitive for all directions $\varphi$ of the perturbing
force. In a similar way, one can show that compass states have
Heisenberg-limited sensitivity to rotation perturbations, that is, the minimal
rotation angle that can be detected is $\theta \simeq s/ |\alpha| \simeq 1/
|\alpha|^2$. To achieve this sensitivity for rotation perturbations, the
circular coherent state has to be first translated in phase space so that the
displaced circle ${\cal C}$ contains the origin of phase space
\cite{Toscano2006}.

%

\begin{figure}
\setlength{\unitlength}{1cm}
\begin{center}
\scalebox{1.2}[1.2]{%
\includegraphics*[width=6.2cm,angle=0]{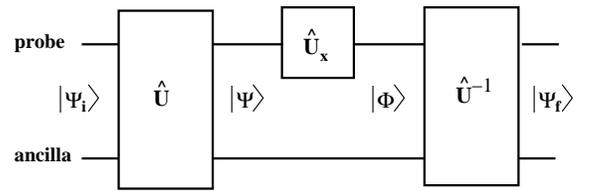}}
\end{center}
\caption{Circuit diagram to measure a small perturbation $\hat{U}_x$ at the
Heisenberg limit. The probe is a harmonic oscillator and the ancilla a two-level system.
The entangling evolution $\hat{U}$ is such that $ | \langle \Psi |
\Phi \rangle|^2 \approx | \langle {\rm cat}_4 | {\rm cat}_4(x) \rangle|^2$, where 
$| {\rm cat}_4 \rangle$ is the compass state defined in Eq.(\ref{defcat4}),
and $| {\rm cat}_4(x) \rangle=\hat{U}_x | {\rm cat}_4 \rangle$ the perturbed compass
state (see text for details).}
\label{fig3}
\end{figure}

Once the direction of the perturbation $\beta$ is known, i.e., the angle $\varphi$), 
the information about its magnitude is encoded in the fidelity function $f(s)$ between 
the unperturbed and the perturbed states of the probe. 
In order to measure the fidelity, a measurement strategy was proposed 
in \cite{Toscano2006}, that is summarized in Fig. (\ref{fig3}).
It involves entangling the probe with an ancillary
system, a two-level system (TLS), and designing their interaction $\hat{U}$ in such a
way that the information about the overlap between the unperturbed and
perturbed probes states can be inferred from measurements of the populations of the
TLS. Starting with an initial joint state $| \Psi_i \rangle=|0\rangle \otimes |g\rangle $,
where $|0\rangle$ is the vacuum state of the probe and $|g\rangle$
the lower state of the ancilla, the evolved states, with and without the perturbation
applied, are $| \Phi \rangle = \hat{U}_x \hat{U} |\Psi_i \rangle$ and $|\Psi
\rangle = \hat{U} | \Psi_i \rangle$ respectively. Here $\hat{U}_x$ denotes
the unitary operator corresponding to the perturbation. The interaction
evolution $\hat{U}$ has to be designed in such a way that $ | \langle \Psi |
\Phi \rangle|^2 \approx | \langle {\rm cat}_4 | {\rm cat}_4(x) \rangle|^2$.
Then, we have to undo the unitary evolution $\hat{U}$ to obtain a final
entangled state  of the composite system 
\beq
\label{final-entangled-state}
| \Psi_f \rangle = 
\sqrt{P_e} |e, \Psi^e_{S} \rangle +
\sqrt{P_g} |g, \Psi^g_{S} \rangle \;\;,
\eeq
where $P_e$ and $P_g=1-P_e$ are the probabilities of measuring the TLS in levels $e$ and $g$, respectively, given by
\begin{equation}
P_g =1-P_e= \frac{ | \langle \Psi_i | \Psi_f \rangle |^2}{ | \langle 0
| \Psi^g_S \rangle|^2}
\approx\frac{ | \langle {\rm cat}_M | {\rm cat}_M(x) \rangle |^2}
{ | \langle 0 | \Psi^g_S \rangle|^2}\;\; .
\end{equation}
As mentioned in \cite{Toscano2006}, this strategy can also be used to measure
the Loschmidt echo, which quantifies the sensitivity of a quantum system to
perturbations. Also, similar hybrid continuous variable-qubit systems have been
proposed for quantum information processing and quantum computation 
\cite{Lee2006,Spiller2006}.

Later in the paper we will show how to create compass states of the
vibrational degree of freedom of a trapped ion, and how to measure weak
perturbations at the Heisenberg limit by entangling the motional degree of
freedom with a two-level system, corresponding to two internal hyperfine
levels of the ion.

\section{Engineering the ion -laser interaction in an ion trap.}

Let us consider a single ion confined inside a harmonic trap. In good
approximation the quantized motion of the ion center-of-mass along each
spatial dimension can be described by a quantum harmonic oscillator. When the
ion is illuminated by laser light quasi-resonant to one of its electronic
transitions its motional degrees-of-freedom can be coupled to the electronic
ones via photon-momentum exchange. The laser excitation can be done in several
different ways, giving rise to a large number of possible interaction
Hamiltonians between electronic and motional degrees-of-freedom. Here we will
be interested in two basic types of laser excitation in a situation where the
motional sidebands can be spectroscopically well resolved. Moreover, we will
consider the excitation of only one motional degree-of-freedom.

\begin{figure}[t]
\setlength{\unitlength}{1cm}
\begin{center}
\scalebox{1.2}[1.2]{%
\includegraphics*[width=6.2cm,angle=0]{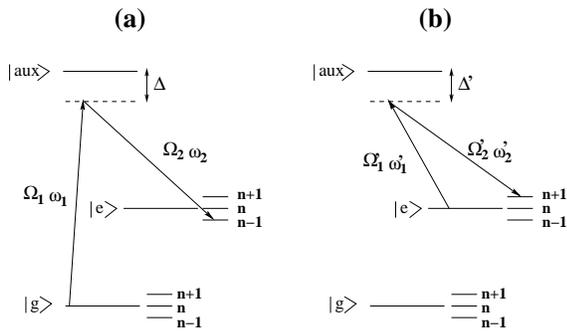}}
\end{center}
\caption{Electronic (internal) and motional (external) energy levels (not to scale)
of a trapped ion, coupled by two Raman lasers. (a) corresponds to the first red sideband
transition between two electronic levels, and (b) corresponds to the first
blue sideband transition via the virtual excitation of an electronic level.}
\label{fig4}
\end{figure}

The first excitation scheme of interest is the Raman excitation of a
dipole-forbidden electronic transition between two hyperfine electronic
states, on resonance to a given motional sideband. This can be done via the
off-resonant excitation of an intermediary electronic level by two laser fields
of adequate frequencies (see Fig.~\ref{fig4}(a)). In this case, the interaction Hamiltonian, in the
interaction picture, is given by~\cite{vo-ma}:
\beq
\label{basic-hamiltonian}
\hat{H}_I=\frac{1}{2}\hbar|\Omega_0|\,e^{i\phi}\,\hat{\sigma}_+
\hat{f}_k(\hat{n},\eta)\;\hat{a}^k +\mbox{H.c.}\;, \eeq where the
operator-valued function $\hat{f}_k(\hat{n},\eta)$ depends on the motional
number operator $\hat{n}\equiv\hat{a}^{\dagger}\hat{a}$, \beq\label{eqfn}
\hat{f}_k(\hat{n},\eta)=e^{-\eta^2/2}\sum_{l=0}^{+\infty}
\frac{(i\eta)^{2l+k}}{l!(l+k)!}\;\frac{\hat{n}!}{(\hat{n}-l)!}\;\;, \eeq and
we define
$\hat{n}!/(\hat{n}-l)!\equiv[\hat{n}-(l-1)][\hat{n}-(l-2)]\ldots\hat{n}$.  The
operators $\hat{\sigma}_+$ and $\hat{a}$ are the electronic flip operator
between the states $|g\rangle$ and $|e\rangle$ and
the annihilation operator of a motional quantum, respectively.
$\Omega_0=|\Omega_0|\,e^{i\phi}$ is the effective Raman Rabi frequency and $k$
corresponds to the excitation of the $k$th lower motional sideband (in the
example of Fig.~\ref{fig4}(a) $k\!=\!1$).  The Lamb-Dicke parameter $\eta$ can
be defined as $\eta=({\bf\delta k_L\cdot u})\Delta x_0$, where $\Delta x_0$ is
the spread of the motional wave function in the ground state of the harmonic
potential and $\bf u$ is a unit vector giving the direction of the motional
degree-of-freedom we are considering. The vector ${\bf \delta k_L}={\bf
k_{L_1}}\!-\!{\bf k_{L_2}}$ is the difference of the wave vectors of the two
Raman lasers. The value of the Lamb-Dicke parameter $\eta$ can be changed via
the geometry of the laser beams in two different ways. The first one consists
in changing the modulus of the vector ${\bf \delta k}$. This modulus reaches
its minimum value when the laser beams are co-propagating and its maximum
value when the laser beams are counter-propagating. The second way consists in
changing the direction of the vector ${\bf \delta k}$ with respect to the unit
vector ${\bf u}$. So, via the laser geometry, it is possible to change the
value of $\eta$ in some extent.  In particular, using co-propagating laser
beams, one can reach the Lamb-Dicke regime where $\eta\ll 1$.

Note that when $\eta\ll 1$ the function $\hat{f}_k(\hat{n},\eta)\approx
(i\eta)^k$. In this case, the excitation of the carrier resonance ($k=0$)
leads to the interaction Hamiltonian: \beq
\label{rotation-hamiltonian}
\hat{H}_I=\frac{1}{2}\hbar|\Omega_0|\,e^{i\phi}\,\hat{\sigma}_+
+\mbox{H.c.}\;. \eeq
This Hamiltonian allows one to perform rotations
of the electronic state of the ion according to the evolution operator 
\beq
\label{theta-pulses}
\hat{U}_{\theta}(\vec{v})\equiv e^{-i\hat{H}_It/\hbar}=
\cos\left(\frac{\theta}{2}\right)-i\sin\left(\frac{\theta}{2}\right)
(v_x\hat{\sigma}_x+v_y\hat{\sigma}_y)\;\;, \eeq where
$\theta=|\Omega_0| t$, $v_x=\cos(\phi)$, $v_y=-\sin(\phi)$ and
$\hat{\sigma}_x$, $\hat{\sigma}_y$ are the Pauli operators.  
Our weak force measurement scheme involves reversing every unitary evolution
applied in the process.  Hence, the inverse evolution $\hat{U}_{\theta}(-\vec{v})$ can be implemented
simply by changing by $\phi\rightarrow\phi+\pi$ the phase of the effective Raman Rabi frequency
$\Omega_0$.

In situations where the value of the parameter $\eta$ is not extremely small
($\eta\approx (0.1,0.2)$) terms of order $\eta^2$ should be taken into account
in Eq.~(\ref{eqfn}).  In this case, for carrier excitation ($k=0$), the
function  $\hat{f}_0(\hat{n},\eta)$ can be written as:
\begin{equation}
\label{approx-fk-order1}
\hat{f}_0(\hat{n},\eta)\approx A_0 + A_1 \hat{n},
\end{equation} 
with $A_0=1-\eta^2/2$ and $A_1=-\eta^2$.  Under this condition, the choice of
$\phi=0$ in the Hamiltonian~(\ref{basic-hamiltonian}) leads to the evolution
operator
\begin{equation}
\label{conditioned-rotation}
\hat{R}_c(\bar{\theta})\equiv e^{-i\hat{H}_It/\hbar}= e^{i\nu\hat{\sigma}_x}
\;e^{i\bar{\theta}\hat{\sigma}_x\hat{n}},
\end{equation}
where $\nu=-|\Omega_0| t A_0/2$ and $\bar{\theta}=-|\Omega_0|t A_1/2$.  The
action of this operator corresponds to a rotation in phase-space of the
motional state of the ion conditioned to its electronic state:  
\bea
\label{action-of-Rc}
\hat{R}_c(\bar{\theta})|\alpha\rangle |\uparrow_x\rangle&=&e^{i\nu}\;
|\alpha\;e^{i\bar{\theta}} \rangle |\uparrow_x\rangle
\nonumber \\
\hat{R}_c(\bar{\theta})|\alpha\rangle |\downarrow_x\rangle&=&e^{-i\nu}\;
|\alpha\;e^{-i\bar{\theta}}\rangle |\downarrow_x\rangle, 
\eea 
where $|\alpha\rangle$ represents a coherent motional state, 
$|\uparrow_x\rangle\equiv (|e\rangle+|g\rangle)/\sqrt{2}$ and
$|\downarrow_x\rangle\equiv (|e\rangle-|g\rangle)/\sqrt{2}$
are eigenstates of the $\hat{\sigma}_x$ Pauli operator, and $|e\rangle$
and $|g\rangle$ represent, respectively, the upper and lower electronic states
of the transition driven by the lasers (eigenstates of the $\hat{\sigma}_z$
Pauli oparator). The inverse operation $\hat{R}_c(-\bar{\theta})$ 
can be implemented choosing $\phi=\pi$.

The dependence of the Hamiltonian~(\ref{basic-hamiltonian}) on the motional
number operator $\hat n$ is determined by the function
$\hat{f}_k(\hat{n},\eta)$. Some control of this dependence is obtained by
changing the value of the Lamb-Dicke parameter $\eta$. However, this control
is rather limited and it is of interest to find ways of extending the
possibilities of tailoring the $\hat n$-dependence of the Hamiltonian
~(\ref{basic-hamiltonian}). As has been shown in \cite{ma-vo1}, some
extra tailoring can be done by exciting the same vibrational sideband of the
ion by $N$ pairs of Raman lasers of arbitrary effective Rabi frequencies
$\Omega_j$ and Lamb-Dicke parameters $\eta_j\,(j\!=\!1,\dots,N)$.  Remember
that the parameter $\eta_j$, corresponding to a given pair of Raman lasers, can
be controlled by varying the geometry of those laser beams.  Under such
excitation scheme the resulting interaction Hamiltonian maintains the general
form of the Hamiltonian~(\ref{basic-hamiltonian}). The combined effect of the
$N$ laser pairs appears in the operator function $\hat{f}_k(\hat{n})$, which
is transformed in the engineered function $\hat{f}_k^{\mbox{\tiny(e)}}(\hat{n})$.  
This function can be written in the form of a Taylor series,
\begin{equation}
\hat{f}_k^{\mbox{\tiny (e)}}(\hat{n})=\sum_{p=0}^\infty\, A_p\, \hat{n}^p,
\label{engineered}
\end{equation}
with the coefficients $A_p$  given by 
\begin{equation}
A_p = \left\{\begin{array} {ccc} \displaystyle\sum_{j=1}^N\,
e^{-\eta_j^2/2}\frac{\Omega_j}{\Omega_L}\frac{(i\eta_j)^k}{k!}
&\mbox{if}& p=0, \\[1ex] (-1)^p \displaystyle\sum_{j=1}^N\,
e^{-\eta_j^2/2}\frac{\Omega_j}{\Omega_L}{(i\eta_j)}^k \alpha_{p,j}
&\mbox{if}& p\neq 0,
\end{array} \right.
\label{Coefficients}
\end{equation}
where  $\Omega_L$ is a reference Rabi
frequency and the coefficients $\alpha_{p,j}$ are definded by: 
\begin{equation}
\alpha_{p,j}=\sum_{m=p}^\infty\,(-1)^{m-p}S_m^{(p)}\frac{\eta_j^{2m}}{m!(m+k)!}\;.
\label{Stirling}
\end{equation}
The $S_m^{(p)}$ are the Stirling numbers of first kind~\cite{abramowitz}, that
count the number of permutations of $m$ elements with $p$ disjoint cycles. The
above expression shows that, for a given set of parameters $\eta_j$, $N$ of
the coefficients $A_p$ can be independently fixed by the values of the $N$
Rabi frequencies $\Omega_j$.  The values of the Rabi frequencies are the
solutions of the set of $N$ linear equations obtained from
Eq.(\ref{Coefficients}) by fixing the values for the $N$ coefficients $A_p$.
The free parameters $\eta_j$ can be used to optimize the coupling between the
electronic and the vibrational degrees of freedom.  This excitation scheme
opens the possibility for engineering a large number of interaction
Hamiltonians.

In the next section we will be interested in engineering the function
\beq
\label{approx-fk-order2}
\hat{f}_k^{\mbox{\tiny (e)}}(\hat{n},\eta)= A_0 + A_1 \hat{n} + A_2\hat{n}^2 +
\mbox{\cal O}(\eta^8_{\mbox{\scriptsize max}}\hat{n}^4),
\eeq
where only the values of $A_2$ and $A_3=0$ have to be fixed independently. This
can be done with $N=2$ pairs of laser beams in Raman configuration. If the two
laser pairs excite the carrier resonance ($k=0$), the resulting interaction
Hamiltonian will be
\beq
\label{kerr-hamiltonian}
\hat{H}_I=\frac{1}{2}\hbar|\Omega_L|\,e^{i\phi}\,\hat{\sigma}_+
\hat{f}_k^{\mbox{\tiny (e)}}(\hat{n},\eta) +\mbox{H.c.}\;. \eeq 
For $\phi=0$ in the above equation one gets the following Kerr-type evolution operator
\beq
\label{kerr-type-operator}
\hat{V}(\phi_2)\equiv e^{-i\hat{H}_It/\hbar}=e^{-i\phi_0\hat{\sigma}_x}
\;e^{-i\phi_1\hat{\sigma}_x\hat{n}}\;e^{-i\phi_2\hat{\sigma}_x\hat{n}^2}\;\;,
\eeq
with $\phi_j=|\Omega_L|A_jt/2$ ($j=0,1,2$).
We index $\hat{V}$ just with $\phi_2$ because, when applying this operation to a trapped ion, 
only the value taken by $\phi_2$  will be of importance. 
Note that the evolution generated by $\hat{V}(\phi_2)$ may be inverted
by setting $\phi=\pi$ in Eq.(\ref{kerr-hamiltonian}). 

The second basic type of laser excitation scheme we are interested in is the
Raman excitation of one motional sideband via the virtual excitation of a
given electronic transition (see Fig.~(\ref{fig4}(b)). In this excitation
scheme only the motional degree-of-freedom is excited conditioned on the
occupation of a specific electronic level. In the example of
Fig.~(\ref{fig4}(b)), if the ion is in the electronic state $|g\rangle$
nothing happens, whereas if the ion is in the electronic state $|e\rangle$ its
vibrational motion will be excited. In this case the interaction Hamiltonian,
describing the action of the lasers on the motional degree-of-freedom, is
given by~\cite{wal-vo,Monroe1996}
\beq
\label{displace0-hamiltonian}
\hat{H}_I=\frac{1}{2}\hbar|\Omega_0|\,e^{i\phi}
\hat{f}_k(\hat{n},\eta)\;\hat{a}^k +\mbox{H.c.}\;. \eeq 
When the first motional sideband is excited ($k=1$) in a situation where
$\eta\ll1$, the above Hamiltonian simplifies to
\beq
\label{displace-hamiltonian}
\hat{H}_I=\frac{1}{2}\hbar|\Omega_0|\,e^{i\phi}\hat{a} +\mbox{H.c.}\;.  
\eeq
The time-evolution generated by the Hamiltonian~(\ref{displace-hamiltonian})
is equivalent to action of the displacement operator $\hat{D}(\alpha)$
\beq
\label{displace-oper} 
\hat{U}(t)\equiv
e^{-i\hat{H}_It/\hbar}=\hat{D}\left(-\frac{1}{2}\eta|\Omega_0|\,e^{-i\phi}t\right).
\eeq 
For this reason, the Hamiltonian~(\ref{displace-hamiltonian}) can be used
to coherently displace the motional state of the ion in phase-space
conditioned on the occupation of a given electronic level. In order to stress
the dependence of the action of the above Hamiltonian on the electronic state
we will represent the evolution operator~(\ref{displace-oper}) by
$\hat{D}_c(\alpha)$.


\section{Dynamical generation of the compass state in ion traps}

According to the general procedure to measure small
perturbations, summarized in Fig.(\ref{fig3}), the 
measurement of weak forces that couple to the motional degree
of freedom of a trapped ion involves the dynamical 
generation of an intermediate maximally entangled state between
the electronic (the ancilla) and the motional (the probe) degree of 
freedom of the ion, i.e.,
\beq
\label{intermediate-state-Psi}
|\Psi\rangle=\frac{1}{\sqrt{2}}\;\;|\mbox{cat}_4\rangle\;|\uparrow\rangle\;
+\frac{1}{\sqrt{2}}\;\;|\overline{\mbox{cat}_4}\rangle\;
|\downarrow \rangle\;\;,
\eeq
where $|\uparrow \rangle$, $|\downarrow \rangle$ are two orthogonal internal 
states of the ion,  and $|\mbox{cat}_4\rangle$, 
$|\overline{\mbox{cat}}_4\rangle$ are compass states (Eq.(\ref{defcat4})) 
of the center-of-mass motion of the ion. 
In the following we describe two ways to dynamically generate
these states starting from an initial state of the form
$| \Psi_f \rangle=|0\rangle \otimes |g\rangle$, where $|0\rangle$  is 
the ground state of the center-of-mass motion of the ion 
and $|g\rangle $ is the lower electronic state considered.
The first approach combines quantum gates over the internal states 
and conditional linear operations over the center-of-mass motion, described in
Section III.
The second approach uses the engineering of a non-linear ion-laser interaction
of the Kerr-type, also described in Section III.


\subsection{First approach}
\label{subsec:first-approach}

\begin{figure}[t]
\setlength{\unitlength}{1cm}
\begin{center}
\scalebox{1.2}[1.2]{%
\includegraphics*[width=7.0cm,angle=0]{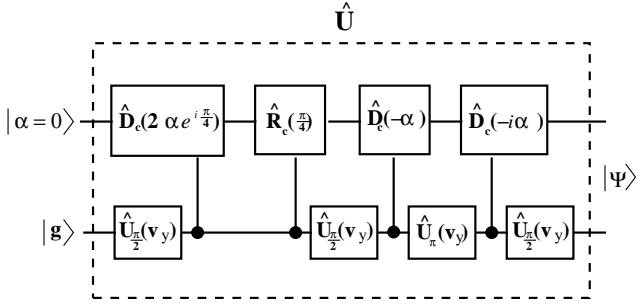}}
\end{center}
\caption{Circuit diagram to generate the compass state via quantum gates. 
Starting with an initial state $| \Psi_i \rangle=|0\rangle \otimes |g\rangle$ of a trapped ion,
where $|0\rangle$ is the ground state of its center-of-mass motion, and $|g\rangle$ is the
lowest of the two hyperfine electronic states considered, this sequence of unitary operations
generates an output state $|\Psi \rangle$ of the form in Eq.(\ref{intermediate-state-Psi}),
with $|\uparrow\rangle\equiv|e\rangle$ and $|\downarrow\rangle\equiv|g\rangle$, and
$|\mbox{cat}_4\rangle$ and $|\overline{\mbox{cat}}_4\rangle$ compass states.
The unitary evolution $\hat{U}$ in Fig.(\ref{fig3}) is composed of this sequence of operations.}
\label{fig5}
\end{figure}
 
In this approach  the set of unitary operations that leads to the state
in Eq.(\ref{intermediate-state-Psi})  is described in Fig. \ref{fig5}.
The first type of unitary operations in the sequence are the quantum gates 
$\hat{U}_{\theta}(\vec{v})$,  given by the carrier $\theta$-pulses in Eq.(\ref{theta-pulses})
that rotate the internal states of the ion around the Bloch vector  $\vec{v}$ by an angle $\theta$.
We only perform rotations around the Bloch vector $\vec{v}=(0,v_y,0)$ that can be
carried out by choosing the phase of the effective Raman Rabi frequency 
$\Omega_0$ equal $\phi=-\pi/2$.   Setting the time duration of the pulse 
$\theta=\pi/2=|\Omega_0|t_{\pi/2}$ we can implement $\hat{U}_{\pi/2}(v_y)$, 
that performs $\pi/2$-rotations, and with $\theta=\pi=|\Omega_0|t_{\pi}$
we implement $\hat{U}_{\pi/2}(v_y)$, that performs $\pi$-rotations. 
The other type of unitary operations in the sequence of Fig.\ref{fig5}
affect the motional state of the ion conditioned on the state of the 
electronic degree of freedom.
$\hat{D}_c(\bar{\alpha})$ given in Eq.(\ref{displace-oper}) displaces a 
coherent state of the vibrational motion of the ion if the internal 
electronic part is in the upper state $|e\rangle$, and does nothing if the
electronic part is in the ground state. That is,  
$\hat{D}_c(\bar{\alpha})|\alpha\rangle|e\rangle=
e^{i {\rm Im}(\bar{\alpha}\alpha^*)} |\alpha+\bar{\alpha}\rangle|e\rangle$,
and $\hat{D}_c(\bar{\alpha})|\alpha\rangle|g\rangle=|\alpha\rangle|g\rangle$.
The conditioned operation $\hat{R}_c(\pi/4)$ in Fig. \ref{fig5}
performs phase space rotations of a coherent state of the 
vibrational motion of the ion according to Eq.(\ref{action-of-Rc}).

The final entangled state  $|\Psi\rangle$ of the sequence of unitary 
operations of Fig. \ref{fig5} is of the form given in Eq.(\ref{intermediate-state-Psi}),
with $|\uparrow\rangle = |e\rangle$ and $|\downarrow\rangle = |g\rangle$,
and the vibrational motion of the ion in compass states of the form
\begin{eqnarray}
|\mbox{cat}_4\rangle &=&
\frac{1}{2} \left(e^{i\nu}|i\alpha\rangle+ e^{-i\nu}|-\alpha\rangle- e^{i\nu}|-i\alpha\rangle+ e^{-i\nu}|\alpha\rangle\right) , \nonumber \\
|\overline{\mbox{cat}}_4\rangle &=&
\frac{1}{2} \left(e^{i\nu}|i\alpha\rangle- e^{-i\nu}|-\alpha\rangle- e^{i\nu}|-i\alpha\rangle-
e^{-i\nu}|\alpha\rangle\right) , \nonumber
\end{eqnarray}
that are equally oriented in phase space.

A realistic estimation for the total time needed to perform the unitary operations in this
approach  can be done using the experimental parameters in \cite{Monroe1996}.
For the $\hat{U}_{\pi/2}(v_y)$ operation the Raman Rabi frequency is of the order 
$\Omega_0/2\pi\approx 250$ kHz, so the $\pi$-pulse duration is about $1\mu$s. 
This implies a total time for single qubit operations of about $2.5 \mu$s.
The conditional displacement $\hat{D}_c(\bar{\alpha})$
of amplitude $|\bar{\alpha}|=\eta\Omega_0 \tau/2$ can be implemented
with a Lamb-Dicke parameter $\eta\approx 0.15$ and an
effective strength of the Raman laser configuration 
$\Omega_0/2\pi\approx 300$kHz. 
Let us suppose that we create compass states with amplitude
$|\alpha|\approx 3$, thus for the maximum displacement in the process
$\bar{\alpha}=2\alpha$ we have a duration time $\tau\approx 21 \mu$s.
Therefore, the total time for the conditional displacements in the process
is aproximately $42\mu$s.
For the duration of the conditional rotation $\hat{R}_c(\bar{\theta})$
we have $\bar{\theta}=\pi/4=\Omega_0\eta^2 t_{\pi/4}/2$, and using also
$\eta\approx 0.15$ and $\Omega_0/2\pi\approx 300$kHz, we get
$t_{\pi/4}\approx 27 \mu$s. Hence, the total time needed to generate the state
$| \Psi \rangle$ is approximately $72 \mu$s.


\subsection{Second approach}
\label{subsec:second-approach}

\begin{figure}[t]
\setlength{\unitlength}{1cm}
\begin{center}
\scalebox{0.95}[0.95]{%
\includegraphics*[width=7.0cm,angle=0]{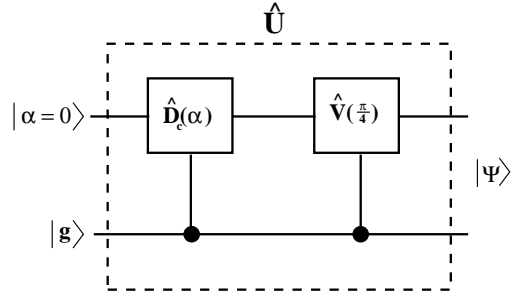}}
\end{center}
\caption{Circuit diagram to generate the compass state via engineered Kerr nonlinearity
$\hat{V}$. The input state is the same as in Fig. \ref{fig5}. The output state $|\Psi \rangle$ has  the form in 
Eq.(\ref{intermediate-state-Psi}), with  $|\uparrow\rangle\equiv | \uparrow_x \rangle$ and 
$|\downarrow\rangle\equiv | \downarrow_x \rangle$, and
$|\mbox{cat}_4\rangle$ and $|\overline{\mbox{cat}}_4\rangle$ compass states.
The unitary evolution $\hat{U}$ in Fig.(\ref{fig3}) is composed of this sequence of operations.}
\label{fig6}
\end{figure}

In this approach  the set of unitary operations that leads to the state
in Eq.(\ref{intermediate-state-Psi})  is described in Fig. \ref{fig6}.  
The displacement operator $\hat{D}_c(\alpha)$ acting on the vibrational
state of the ion can be implemented along the lines of Section III, where now the 
Raman excitation of one motional sideband is via the virtual excitation of the
electronic transition with the lower state $|g\rangle$.
The engineered Kerr operation
$\hat{V} = e^{-i \phi_0 \hat{\sigma}_x} e^{-i \phi_1 \hat{\sigma}_x \hat{n}} 
e^{-i \phi_2 \hat{\sigma}_x \hat{n}^2}$  generates all the circular coherent
states in Eq.(\ref{defcatM}) of the vibrational degree
of freedom for $\phi_2 = \pi/M$ \cite{compass-state-generation}. 
In particular, we can see that the compass state ($M=4$) is generated  
because $e^{\pm i(\pi/4)\hat{n}^2}|\alpha\rangle = \frac{1}{2} \left( e^{\pm i(\pi/4)}
\left( |\alpha\rangle-|-\alpha\rangle \right)+ \left( |i\alpha\rangle + |-i\alpha\rangle \right)\right)$. 
The total unitary operation is
$\hat{U} = \hat{V}(\pi/4) \hat{D}_c(\alpha)$, and
the output state $|\Psi \rangle$ is of the form given in Eq.(\ref{intermediate-state-Psi}),
with the electronic part as 
$|\uparrow \rangle = |\uparrow_x \rangle$ and $|\downarrow \rangle = |\downarrow_x \rangle$,
and the vibrational motion of the ion in compass states of the form
\begin{eqnarray}
|\mbox{cat}_4\rangle &=& \frac{e^{-i \phi_0}}{2} 
\left[
e^{-i \pi /4} \big( |\bar{\alpha} \rangle - |-\bar{\alpha} \rangle \big) +
| i \bar{\alpha} \rangle + | -i \bar{\alpha} \rangle \right], \nonumber \\
|\overline{\mbox{cat}}_4\rangle &=&
- \frac{e^{i \phi_0}}{2} 
\left[
e^{i \pi /4} \big( |\tilde{\alpha} \rangle - |-\tilde{\alpha} \rangle\big) +
| i \tilde{\alpha} \rangle + | -i \tilde{\alpha} \rangle \right].
\nonumber
\end{eqnarray}
Here $\tilde{\alpha} = \alpha e^{i \phi_1}$, and $\bar{\alpha}=\alpha e^{-i \phi_1}$.
These two compass states are rotated with respect 
to each other in an angle $2 \phi_1$. 
If it is desired to have the compass states equally oriented in phase space, 
as in the first approach, a conditional rotation could be performed.

The engineering of the interaction Hamiltonian can be accomplished with $N=2$ pairs
of Raman lasers (with Lamb-Dicke parameters $\eta_j$ and Rabi frequencies $\Omega_j$,
$j=1,2$) driving resonantly the carrier ($k=0$) transition. Indeed, one can choose the
coefficient of the quadratic $\hat{n}^2$ part in Eq.(\ref{approx-fk-order2}) to be $A_2=1$, 
and impose that the cubic coefficient identically vanishes, $A_3=0$.  
This sets the time $t^*$ when the
compass state is generated as  $\phi_2= |\Omega_L | A_2 t^*/2 = \pi/4$.
The compass state is formed at the time $t^*$ irrespective of the values that the phases
$\phi_0$ and $\phi_1$ take. Thus, we do not require any fixed values for $A_0$ and $A_1$
in the interaction Hamiltonian, simplifying the engineering design to only two pairs of Raman lasers. 
Selecting the set of Lamb-Dicke parameters $\eta_j$ for the two pair of lasers as $\{0.4,0.35\}$ 
yields for the relative Rabi frequencies $\Omega_j/\Omega_L$ the values
$\{-520, 1154\}$. This requires that the Rabi frequencies of the two pair of lasers be related
as $|\Omega_2| \approx 2.2 |\Omega_1|$, and have a $\pi$ phase shift with respect to each other.
The coefficients $A_0$ and $A_1$ are dependent quantities that can be obtained from
Eqs.(\ref{Coefficients},\ref{Stirling}),  and are equal to $A_0 \approx 605$ and $A_1 \approx -57$.
In order to minimize $t^*$ we choose the highest experimentally available Rabi frequencies,
$|\Omega_1| \approx 5$MHz and $|\Omega_2| \approx 11$MHz. Therefore,
the reference Rabi frequency $\Omega_L$ is $|\Omega_L| \approx 9.5 \times 10^3 {\rm s}^{-1}$, and
the time $t^*$ at which the compass states is generated is equal to 
$t^*= \pi/ 2 |\Omega_L| \approx 165 \mu{\rm s}$. The total time needed to generate the final state $|\Psi \rangle$ in Fig.6
is equal to the duration of the conditional displacement (approximately equal to $10 \mu{\rm s}$
for $|\alpha| \approx 3$, see previous sub section), plus the time $t^*$ for the compass state
generation, that is a total time of approximately $175 \mu{\rm s}$.


\section{Weak force detection scheme}

Once the intermediate state of the form Eq.(\ref{intermediate-state-Psi}) 
is created, either  by means of the first or second approaches, the 
probe is subjected to the perturbation to be measured. 
We are interested in weak classical forces that couple
to the center-of-mass motion of the ion, and thus cause a small linear displacement
of the motional quantum state of the ion, irrespective of its internal electronic
state. This can be simulated by applying the displacement operator
$\hat{D}_c(\beta=e^{i\varphi}s\alpha/|\alpha|$) on the vibrational
state of the ion with two pairs of Raman lasers, each of them inducing
an equal blue sideband transition on the  $|g\rangle$ and $|e\rangle$ states
(see Section III). Alternatively, a uniform electric field oscillating near the trap frequency
results in a displacement perturbation acting equally on both $|g\rangle$ and $|e\rangle$
\cite{Myatt2000}.  The general form of the created perturbed state is
\begin{equation}
|\Phi\rangle=\frac{1}{\sqrt{2}}\;\;|\mbox{cat}_4(s)\rangle\;|\uparrow\rangle\;
+\frac{1}{\sqrt{2}}\;\;|\overline{\mbox{cat}_4(s)}\rangle\;
|\downarrow \rangle\;\;.
\end{equation}

After reversing the evolution along the lines described in Sections III and 
IV we get the final entangled state of the form in  Eq.(\ref{final-entangled-state}). 
In the first approach we have
\beq
\sqrt{P_g} |g, \Psi^g_{S} \rangle  =\left(
 A_1\;|0\rangle+A_2\;
|-\bar{\alpha}\rangle+
A_2^*\;
|\bar{\alpha}\rangle\right) \otimes |g\rangle \;\;,
\eeq
where $\bar{\alpha}\equiv 2\alpha e^{i\frac{\pi}{4}}$ and
$A_1 = A_1\big(|\alpha|s\big)\equiv \cos\big(b_{+}(\varphi)|\alpha|s\big)\times$
$\times\cos\big(b_{-}(\varphi)|\alpha|s\big)$,
$A_2 = A_2(|\alpha|s)\equiv (e^{-i2\sin(\varphi)|\alpha|s}-e^{i2\cos(\varphi)|\alpha|s})/4$,
with $b_{\pm}(\varphi)\equiv \cos(\varphi)\pm\sin(\varphi)$.
Hence, neglecting contributions of the order
${\cal O}(e^{-|\alpha|^2})$ the probability to measure the ion in the internal state $|g\rangle$ is given by
\beq
\label{Pg-first-approach}
P_g(s)=A_1^2\big(|\alpha|s\big)
\left(1+2\frac{\left|A_2\big(|\alpha|s\big)\right|^2}{A_1^2\big(|\alpha|s\big)}\right)\;\;,
\eeq
where we recognize in $A_1^2\big(|\alpha|s\big)$ the approximation in 
Eq.(\ref{fidelity2-cat4}) for the fidelity function 
$f(s) \equiv|\langle \mbox{cat}_4|\mbox{cat}_4(s)\rangle|^2$.
Thus, $P_g(s)$ exhibits the characteristic oscillatory
behavior with a typical wavelength $\sim1/|\alpha|$ that allows,
after inverting the relation in Eq.(\ref{Pg-first-approach}), the measurement
of $s$ with Heisenberg limit precision. 

In order to invert the function $P_g(s)$ to obtain the small displacement
$s$ we have to know the prior information that $0\leq s \leq s_0$, 
where $s_0$, given in Eq.(\ref{def-so}), is the first zero of $P_g(s)$.
This is not a restrictive condition since we have the flexibility of setting up 
the value of $|\alpha|$ in the experiment in order for $s_0=\pi/2b(\varphi)|\alpha|$
to be an upper bound of the expected displacement.
A good estimation of the unknown parameter $s$ requires repeating the measurement
several times. 
The uncertainty in the determination of the true parameter $s$ can be estimated
if we observe that after $R$ repetitions the probability that the outcome $|g\rangle$
is obtained $r$ times is given by a binomial distribution, that in the large-$R$
limit can be approximated by the Gaussian distribution in the variable
$\xi\equiv r/R$, which can be regarded as effectively continuous \cite{Toscano2006,Luis2004}.
In this limit the probability distribution for the estimator
$\tilde{s}=P^{-1}_g(\xi\equiv r/R)$ is
\begin{equation}
P(\tilde{s}) \approx  \frac{1}{\sqrt{2 \pi \Delta \tilde{s}^2}} \; 
e^{ - (\tilde{s} -s )^2/2 \Delta \tilde{s}^2} ,
\end{equation}
where the uncertainty of $\tilde{s}$ is 
\beq
\Delta \tilde{s}=\frac{\sqrt{(1-P_g)P_g}}{\sqrt{R}|\alpha|}
\left|\frac{\partial P_g}{\partial y}\right|^{-1}_{\tilde{y}=y} ,
\eeq
where $P_g\equiv P_g(s)$ and $y\equiv |\alpha|s$.
Hence, we see that we reach the Heisenberg precision for displacement since
$\sqrt{R}|\alpha|\approx \sqrt{R \bar{n}}$ and
$R \bar{n}$ is the total number of photons used in the measurement.

In the second approach we have
\begin{eqnarray}
\sqrt{P_g} |g, \Psi^g_{S}\rangle  &=& \left(B_1\,|0\rangle+B_2\,|-2\alpha\rangle+
B_3\,|(i-1)\alpha\rangle\right.
\nonumber\\
&&\left.
+B_4\,|-(i+1)\alpha\rangle\right) \otimes |g\rangle\,, 
\end{eqnarray}
where 
\bea
B_1 &\equiv&
(\cos(a_s^{+}|\alpha|s)+\cos(a_c^{+}|\alpha|s)+ \nonumber \\
&&+\cos(a_s^{-}|\alpha|s)+\cos(a_c^{-}|\alpha|s))/2 \;, \nonumber \\
B_2 &\equiv&
(\cos(a_s^{+}|\alpha|s)-\cos(a_c^{+}|\alpha|s)+ \nonumber \\
&&+\cos(a_s^{-}|\alpha|s)-\cos(a_c^{-}|\alpha|s))/2\;, \nonumber \\
B_3 &\equiv&
i(e^{-i(\frac{\pi}{4}+|\alpha|^2)}(\sin(a_s^{+}|\alpha|s)+\sin(a_c^{-}|\alpha|s))+ \nonumber \\
&&+e^{i(\frac{\pi}{4}-|\alpha|^2)}(\sin(a_s^{-}|\alpha|s)+\sin(a_c^{+}|\alpha|s)))/2 \;, \nonumber \\
B_4 &\equiv&
i(e^{-i(\frac{\pi}{4}-|\alpha|^2)}(\sin(a_s^{+}|\alpha|s)-\sin(a_c^{-}|\alpha|s))+ \nonumber \\
&&+e^{i(\frac{\pi}{4}+|\alpha|^2)}(\sin(a_s^{-}|\alpha|s)-\sin(a_c^{+}|\alpha|s)))/2 \;, \nonumber
\eea
and we define 
$a_s^{\pm}\equiv 2\sin(\varphi\pm\phi_1)$,
$a_c^{\pm}\equiv 2\cos(\varphi\pm\phi_1)$.
In this case, the probability of measuring the ion in the internal state
$|g\rangle$ is
\bea
P_g(s)&=&\frac{1}{2}
[1+
(\cos(a_s^{+}|\alpha|s)+\cos(a_c^{-}|\alpha|s)+ \nonumber \\
&&+\cos(a_s^{-}|\alpha|s)\cos(a_c^{+}|\alpha|s))/2]\;,
\eea
that also oscillates with a typical wavelength $\sim 1/|\alpha|$.
Hence, following the steps described for the first approach we
see that after inverting the function $P_g(s)$ we get also in this case
the displacement $s$ with Heisenberg limit precision.


\section{Discussion}

The scheme proposed in this paper for Heisenberg-limited sensitivity to
perturbations with continuous variables relies on the creation and
manipulation of mesoscopic superpositions states of the motional degree of
freedom of a trapped ion, that possesses sub-Planck phase-space structures.
Any decoherence process, such as amplitude or phase decoherence affecting such
quantum superpositions \cite{Myatt2000}, may destroy such small scale
structures, limiting the usefulness of the method for quantum metrology
applications. Typical damping times of the vibrational degree of freedom of a
trapped ion are of the order of $100$ms (see ref.~\cite{deslauriers} and
references therein), so that the typical decoherence time
for the vibrational compass state with $|\alpha| \approx 3$ would be od the
order of $10$ms. This decoherence time-scale should be much larger than the
total interaction time for weak force detection (generation of the compass
state, application of the perturbation, and inversion of the dynamics),
whether the first or the second approaches for compass state generation is
used.  Assuming that the duration of the displacement perturbation takes
approximately $3 \mu{\rm s}$ (compatible with the perturbations used in
\cite{Myatt2000} for engineering the ion's reservoir), the total interaction
time using the first approach is around $150 \mu{\rm s}$, and using the second
approach $353 \mu{\rm s}$. We see that the typical decoherence times are much
larger than the total interaction times in both approaches, pointing towards
the experimental viability of the proposed scheme for quantum-enhanced
measurements using the motional state of a trapped ion.

The total interaction time for the compass state ($M=4$) generation  is shorter when the 
first approach is implemented. It is worth noting, however, that the same tailored  Hamiltonian 
implemented with the second approach generated higher order ($M>4$) circular coherent  states for shorter  times, as the necessary generation time scales as $\phi_2 = \pi/M$. Such higher  order superpositions of coherent states also have sub-Planck phase-space structures, and in principle  could also be used  for Heisenberg-limited quantum metrology.

\acknowledgements

We are grateful to Dana Berkeland, John Chiaverini, Luiz Davidovich and Wojciech H. Zurek for discussions.   FT and RMF acknowledge the support of the program Millennium 
Institute for Quantum Information and of the Brazilian agencies FAPERJ and CNPq. FT thanks Los Alamos  National Laboratory for the hospitality during his stay.


\end{document}